\documentclass[10pt,twocolumn,letterpaper]{article}

\usepackage[pagenumbers]{cvpr} 

\usepackage{graphicx}
\usepackage{amsmath}
\usepackage{amssymb}
\usepackage{booktabs}
\usepackage{longtable}
\usepackage{times}
\usepackage{epsfig}
\usepackage{multirow}
\usepackage{subcaption}
\usepackage[table,xcdraw]{xcolor}
\usepackage[nolist]{acronym}

\usepackage[pagebackref,breaklinks,colorlinks]{hyperref}

\usepackage[capitalize]{cleveref}
\crefname{section}{Sec.}{Secs.}
\Crefname{section}{Section}{Sections}
\Crefname{table}{Table}{Tables}
\crefname{table}{Tab.}{Tabs.}

\def\cvprPaperID{5225} 
\def\confName{CVPR}
\def\confYear{2023}

\begin{document}
	
	\title{HyperKon: A Self-Supervised Contrastive Network for Hyperspectral Image Analysis}
	
	\author{Daniel L Ayuba \and Belen Marti-Cardona \and Jean-Yves Guillemaut\\ \and Oscar Mendez Maldonado\\
		University Of Surrey\\
		{\tt\small \{d.ayuba, b.marti-cardona, j.guillemaut, o.mendez\}@surrey.ac.uk}
}

\maketitle

\begin{abstract}
	The exceptional spectral resolution of hyperspectral imagery enables material insights that are not possible with RGB or multispectral images. Yet, the full potential of this data is often underutilized by deep learning techniques due to the scarcity of hyperspectral-native CNN backbones. To bridge this gap, we introduce HyperKon, a self-supervised contrastive learning network designed and trained on hyperspectral data from the EnMAP Hyperspectral Satellite\cite{kaufmann2012environmental}. HyperKon uniquely leverages the high spectral continuity, range, and resolution of hyperspectral data through a spectral attention mechanism and specialized convolutional layers. We also perform a thorough ablation study on different kinds of layers, showing their performance in understanding hyperspectral layers. It achieves an outstanding 98\% Top-1 retrieval accuracy and outperforms traditional RGB-trained backbones in hyperspectral pan-sharpening tasks. Additionally, in hyperspectral image classification, HyperKon surpasses state-of-the-art methods, indicating a paradigm shift in hyperspectral image analysis and underscoring the importance of hyperspectral-native backbones.
\end{abstract}
\vspace{-5mm}
\section{Introduction}
\label{sec:intro}
\acp{hsi}, with their ability to capture detailed spectral information across hundreds of contiguous bands, have rapidly advanced the capabilities of remote sensing analysis in various domains, including agriculture, mineralogy, and environmental monitoring \cite{cheng2019prospect, brisco1998precision}. This high-dimensional data offers a rich representation of scenes, enabling finer material distinctions than traditional RGB and multispectral images \cite{da2022hyperspectral}. However, the exploitation of \ac{hsi}, particularly using deep learning techniques initially designed for RGB images, presents considerable challenges \cite{yu2017convolutional}.

\acp{cnn} which dominate the computer vision landscape, are predominantly trained and evaluated on RGB data \cite{signoroni2019deep}. These models often struggle to generalize effectively to hyperspectral data due to the vast difference in spectral resolution and the unique characteristics of hyperspectral images \cite{shi2022hyperspectral, bouchoucha2023robustness, signoroni2019deep}. Moreover, the scarcity of native hyperspectral backbones necessitates extensive fine-tuning for their application in this domain.

While Transformers have shown promise in various computer vision tasks \cite{dosovitskiy2020image, zhang2021multi, cai2022semi}, their application to hyperspectral data is not without challenges. Transformers, which are primarily designed for global feature extraction, may not adequately capture the local spectral-spatial nuances fundamental in hyperspectral images, and their high computational requirements and the requirement for a large amount of training data frequently make them less feasible for hyperspectral applications, particularly in scenarios with limited data availability \cite{hong2021spectralformer}.

Deep learning, leveraging both spatial and spectral data, has shown exceptional performance in hyperspectral image tasks such as classification \cite{yu2017convolutional}, segmentation \cite{liu2023multi}, object detection \cite{chhapariya2022cnn}, and unmixing \cite{wang2023ssanet}. Despite this, the reliance on RGB-trained backbones in hyperspectral research implies a lack of full exploitation of the spectral potential, often leading to domain adaptation issues \cite{rangnekar2022learning}.

To address these limitations and the scarcity of hyperspectral-specific solutions, we introduce HyperKon, a novel self-supervised contrastive network trained exclusively on \acp{hsi} from \ac{enmap}. HyperKon, distinct from generic \ac{cnn} backbones, is designed to fully harness the unique properties of hyperspectral data, presenting an innovative approach in hyperspectral image analysis.

\noindent The main contributions of this paper are:
\begin{enumerate}
    \item HyperKon: a hyperspectral-native \ac{cnn} backbone that can learn useful representations from large amounts of unlabelled data.
    \item EnHyperSet-1: An \ac{enmap} dataset curated for use in precision agriculture and other deep learning projects.
    \item HyperSpectral Perceptual Loss: A perceptual loss function emphasizing minimizing errors in the spectral domain.
    \item Demonstration that the representations learned by HyperKon improve performance in hyperspectral downstream tasks.
\end{enumerate}

\section{Related Work}
\label{sec:related_work}
\subsection{Hyperspectral Sensors}

Recent advances in hyperspectral sensors stem from the rising demand for intricate spectral details in areas like remote sensing, environmental monitoring, and sustainable agriculture. These sensors allow for in-depth examination of the chemical and physical characteristics of objects, especially important in agriculture where reflected radiation from plants is important.

The \ac{enmap} project by the \ac{dlr}, introduced in 2022, boasts a cutting-edge hyperspectral sensor catering to diverse applications, from agriculture to mineral exploration \cite{kaufmann2012environmental}. Covering wavelengths from 420 nm to 2450 nm, its imager delivers high-resolution spectral data with pixel sizes of 30m by 30m of the Earth's surface. Public access to \ac{enmap} data has paved the way for AI-driven in-depth analyses.

Anticipated satellite missions, like the \ac{chime} by the \ac{esa} scheduled for 2028, aim to elevate hyperspectral imaging by capturing over 400 spectral bands \cite{nieke2018towards}. These advancements highlight the growing need for advanced computer vision techniques tailored to hyperspectral data.

\subsection{Deep Learning for Hyperspectral Analysis}
Deep learning, particularly through the use of \acp{cnn} and attention mechanisms like Transformers, has significantly advanced \ac{hsi} applications. These techniques have improved performance in classification \cite{li2019deep}, segmentation \cite{minaee2021image}, denoising \cite{tian2020deep}, and spatial resolution enhancement.

\textbf{\ac{hsi} Classification and Segmentation:} Early works like \cite{hu2015deep} demonstrated the superior accuracy of \acp{cnn} over traditional algorithms in \ac{hsi} classification. Innovations in network architectures, such as those introduced by \cite{li2018deepunet}, incorporating global pooling layers and channel attention mechanisms, have furthered this progress, yielding competitive results on benchmarks \cite{perazzi2016benchmark, gupta2019lvis}.

\textbf{Enhancing Spatial Resolution:} The inherent challenge of low spatial resolution in hyperspectral images, due to constraints in spaceborne sensors \cite{adao2017hyperspectral}, has led to advancements in \ac{hsr} and pan-sharpening techniques. Techniques ranging from ResNet-based approaches \cite{han2019deep} to multiscale \acp{cnn} \cite{hu2020pan} and combinations of variational autoencoders with deep residual networks \cite{chen2021hyperspectral} have been developed. These advancements are crucial for improving the spatial resolution of \ac{hsi}, thereby enhancing the performance of downstream tasks.\\
\textbf{The Need for Hyperspectral-Native Solutions:} However, a notable gap in the field is the reliance on RGB-trained backbones for these methods \cite{zhang2021survey, loncan2015hyperspectral}. Such backbones may not fully capture the unique spectral details crucial in hyperspectral data, especially in applications requiring fine spectral discrimination.
\vspace{-2mm}
\subsection{RGB-Backbones vs. Hyperspectral-Native Approaches in Deep Learning}

\textbf{The Role of \acp{cnn} and Transformers:} \acp{cnn} have been foundational in image analysis, demonstrating their capability in tasks like classification \cite{li2019deep, affonso2017deep}, object detection \cite{zhao2019object}, and more \cite{voulodimos2018deep}. The success of \acp{cnn} is largely attributed to their robust feature extraction backbones, as seen in models like VGG \cite{simonyan2014very}, ResNet \cite{he2016deep}, and DenseNet \cite{iandola2014densenet}. Parallel to \acp{cnn}, attention mechanisms such as Transformers have gained traction in computer vision, exemplified by the \ac{vit} \cite{dosovitskiy2020image}, which offers comparable performance to \ac{cnn}-based models in image recognition.\\
\textbf{Transformers in Hyperspectral Classification:} Transformers have shown promise in hyperspectral data classification, capturing global spectral-spatial features effectively. Examples include the spatial-spectral transformer network \cite{he2021spatial} and other models utilizing self-attention mechanisms for hyperspectral classification \cite{gao2021deep, huang20223}.\\
\textbf{Our Focus and Contribution:} Despite the advancements in both \acp{cnn} and Transformers, our research is oriented towards leveraging the strengths of \acp{cnn} for developing hyperspectral-native solutions. We chose \acp{cnn} due to their proven effectiveness in processing and recognizing complex spatial hierarchies in multi-dimensional data, a key requirement for hyperspectral image analysis \cite{hong2021spectralformer}. Furthermore, the parameter efficiency and adaptability of \acp{cnn} make them well-suited for handling the vast spectral diversity inherent in hyperspectral data \cite{yu2017convolutional}. In this context, we propose HyperKon, a \ac{cnn}-based network specifically designed to harness the rich spectral information in hyperspectral images. HyperKon aims to address the critical gap in hyperspectral-native deep learning architectures, offering enhanced capabilities for applications like precision agriculture, where recognizing subtle spectral differences is crucial.
\begin{figure*}[hbt!]
	\centering
	\includegraphics[width=0.8\textwidth]{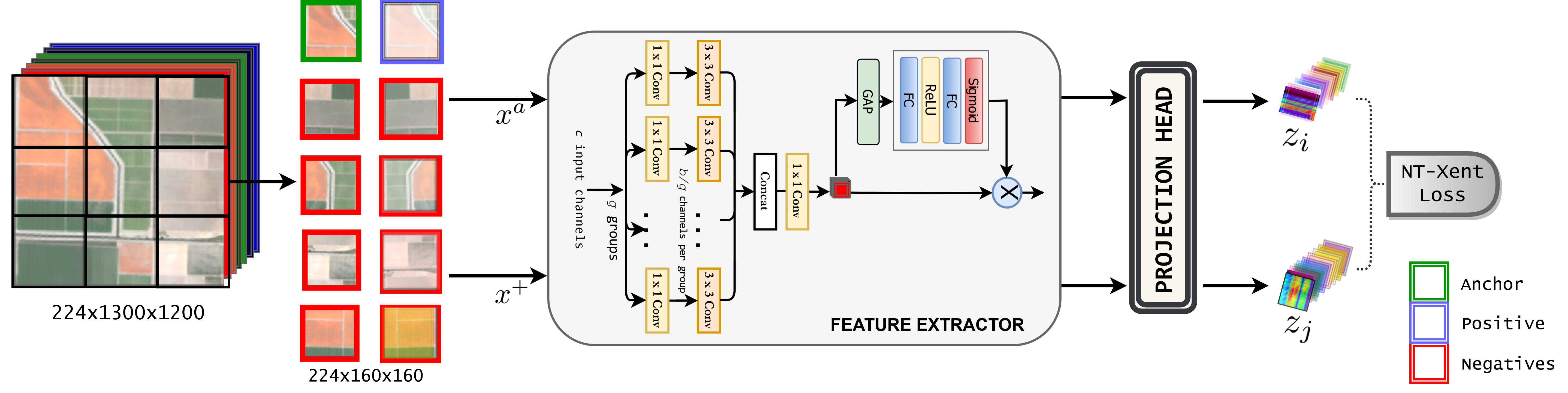}
	\caption{General Overview of HyperKon System Architecture}
	\label{fig:HyperKon_Architecture}
\end{figure*}

\section{Methodology}
\subsection{Hyperspectral Backbone Architecture}
\begin{figure}[hbt!]
	\centering
	\includegraphics[width=0.34\textwidth]{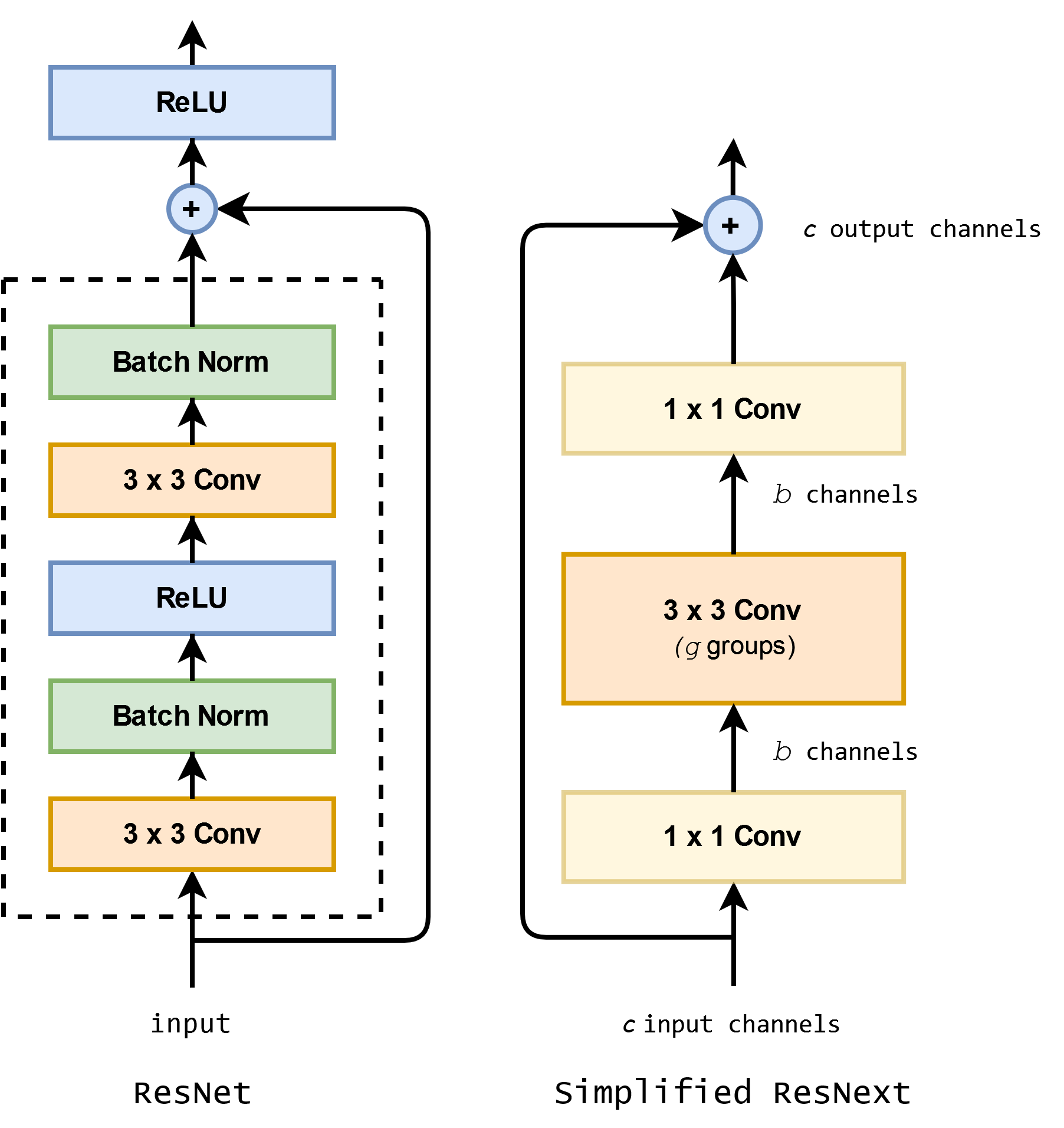}
	\caption{Comparison of ResNet and ResNeXt architectures}
	\label{fig:resnet_vs_resnext}
\end{figure}

The HyperKon network architecture as shown in Figure \ref{fig:HyperKon_Architecture} is designed to effectively handle the complexity and high dimensionality inherent in hyperspectral data. Influenced by the principle of multibranch cardinality, the network is configured into multiple parallel paths, optimizing representational power and computational efficiency. This technique, central to the ResNeXt architecture \cite{wu2020three}, strikes a balance between enhancing representational capability and ensuring computational efficiency, particularly beneficial for data with substantial dimensional depth, like hyperspectral images. Furthermore, as depicted in Figure \ref{fig:resnet_vs_resnext}, the ResNeXt architecture has fewer parameters, making it have reduced sensitivity to learning rates and other hyper-parameters, especially when compared to its predecessor, ResNet\cite{he2016deep}.
\vspace{-1.3mm}
Developing an effective feature extractor for this network requires overcoming two main challenges: capturing the complex, high-dimensional spectral, and spatial features of hyperspectral data and addressing the computational constraints associated with such high-dimensional data.
\begin{figure}[hbt!]
	\centering
	\includegraphics[width=0.3\textwidth]{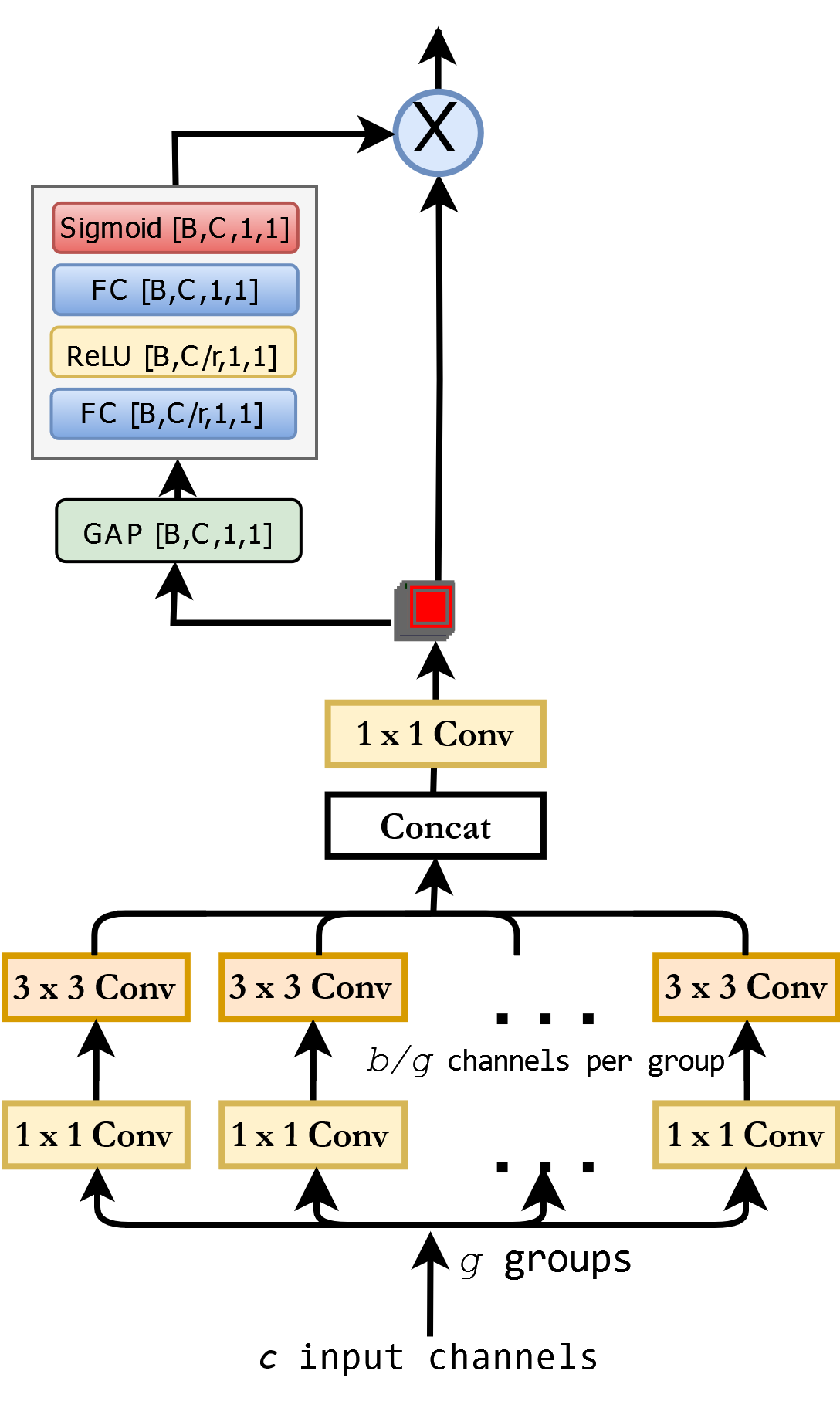}
	\caption{The Feature Extractor used in HyperKon}
	\label{fig:HyperKon_SEB}
\end{figure}

\noindent In the case of Hyperspectral imagery, it is desirable to recalibrate feature maps so that they model channel interdependencies. In our architecture, this is achieved by a novel SEB-based architecture, shown in figure \ref{fig:HyperKon_SEB}. Given an input feature map \(X \in \mathbb{R}^{H \times W \times C}\), the squeeze operation computes channel-wise statistics. It enhances the network's ability to distill and prioritize spectral and spatial information from the high-dimensional data, thus ensuring that the computational resources are optimally utilized.
\noindent The excitation operation then uses a gating mechanism:
\begin{equation}
		\begin{aligned}
			s_c &= \frac{1}{H \times W} \sum_{i=1}^{H} \sum_{j=1}^{W} x_{ijc}, \\
			e_c &= \sigma(\beta \cdot s_c + \gamma), 
		\end{aligned}
\end{equation}
\noindent where \(\sigma\) is the sigmoid activation, and \(\beta\) and \(\gamma\) are trainable parameters. The recalibrated feature map \(X' \in \mathbb{R}^{H \times W \times C}\) is given by:
\begin{equation}
	x'_{ijc} = e_c \cdot x_{ijc}
\end{equation}
\noindent This recalibration emphasizes channels with high interdependencies and suppresses others.
The feature extractor provides preliminary feature representations which, while powerful, might not be fine-tuned to the specific challenges presented by hyperspectral data. To address this, we introduce a projection head. Given the deep feature representation \(F \in \mathbb{R}^{D}\) from the feature extractor, the projection head transforms it into a new space \(Z \in \mathbb{R}^{D'}\):
\begin{equation}
	Z = W_2 \sigma(W_1 F + b_1) + b_2
\end{equation}
\noindent where \(W_1 \in \mathbb{R}^{D' \times D}\), \(W_2 \in \mathbb{R}^{D' \times D'}\), \(b_1 \in \mathbb{R}^{D'}\), and \(b_2 \in \mathbb{R}^{D'}\) are trainable parameters, and \(\sigma\) is a non-linear activation function. This transformation aids in emphasizing discriminative features for hyperspectral data.
Having established the architectural foundation, we now focus on the learning approach that leverages the power of this architecture.

\subsection{Contrastive Learning}
The refined model uses the NT-Xent\cite{chen2020simple} loss function, enhancing its ability to learn distinct features and proving valuable for various hyperspectral image analyses.

\subsubsection{Self-supervised contrastive loss:}
The complexity of hyperspectral data demanded a versatile loss function capable of handling its high dimensionality and suitable for smaller batch sizes. While Triplet loss\cite{schroff2015facenet} and Info-NCE\cite{oord2018representation} can be effective, they require meticulous sample selection, especially challenging for hyperspectral data due to its spectral domain variance. NT-Xent loss\cite{chen2020simple}, however, is adaptable to varying batch sizes, uses cosine similarity for feature vector orientation, and performs well with high-dimensional data, making it the ideal choice for HyperKon. The NT-Xent loss is defined as:
\begin{equation}
	{\ell}_{(z_i, z_j)} = -\frac{1}{N}\sum_{i=1}^{N} \log \frac{\exp(sim(z_i,z_{i'} )/\tau)}{\sum_{j=1}^{2N} \exp(sim(z_i,z_j)/\tau)}
\end{equation}
where $N$ is the number of samples, $z_i$ is the feature representation of sample $i$, $sim(z_i,z_j)=z_i^Tz_j$ is the cosine similarity between $z_i$ and $z_j$, $\tau$ is a temperature parameter that controls the smoothness of the distribution, and $z_{i'}$ refers to the positive feature representation for the query sample $i$.

\subsubsection{HSI Contrastive Sampling:}
Self-supervised contrastive learning heavily leans on the use of image augmentation to extract valuable features from the input data \cite{purushwalkam2020demystifying}. Yet, standard augmentation techniques commonly employed for RGB images \cite{jaiswal2020survey} don't have a specific equivalent explicitly designed for \ac{hsi}.
Let's denote $x_i \in I_k^h$ as a sequence of patches $(x_i)$ taken from a series of hyperspectral images $(I_k^h \forall k=0...S)$ where $S$ represents the dataset size. These patches can then be subjected to a sequence of transformations $A={a_1, a_2...a_n}$, which are enacted on the patches $x_i$ as follows:
\begin{equation}
	x_i' = a_n(...a_2(a_1(x_i))...)
\end{equation}
During the selection of triplet pairs, the hard negative pairs method \cite{robinson2020contrastive} is implemented to pick negative pairs within the same batch. This method aids in choosing negative samples that pose a difficulty to distinguish from the positive samples, thereby enhancing the contrastive learning experience.

The mathematical formulation for selecting hard negative pairs within a batch is as follows: \\
Let's denote $f$ as the encoder function that takes an input patch $x$ and produces the corresponding embedding $z = f(x)$. Provided a batch of $B$ patches ${x_1, x_2, ..., x_B}$, the embeddings ${z_1, z_2, ..., z_B}$ can be computed using the encoder function. The cosine similarity between the embeddings $z_i$ and $z_j$ is measured by the similarity function $s(z_i, z_j)$, which is used to select the hard negative pairs. 
Therefore, for a specific anchor patch $x^a$ and its positive pair $x^+$, a hard negative pair $x^-$ is sampled by:
\begin{equation}
	x^- = \underset{x_k\in\{x_1, x_2, ..., x_B\}\backslash\{x^a, x^+\}}{\text{argmax}}\ s(f(x^a), f(x_k))
\end{equation}
\begin{figure}[hbt!]
	\centering
	\includegraphics[width=0.4\textwidth]{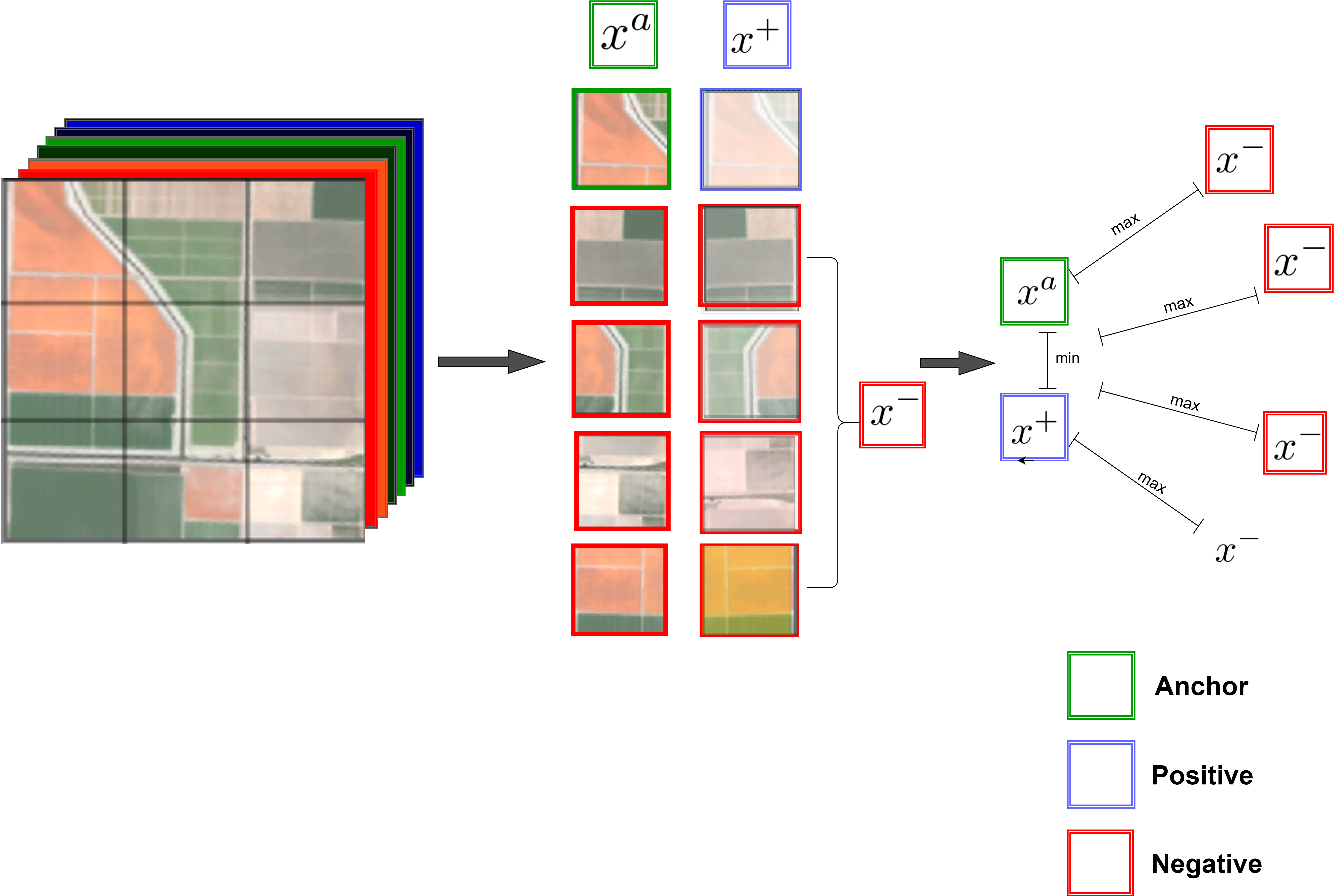}
	\caption{Illustration of HSI Contrastive sampling}
	\label{fig:sampling}
\end{figure}
This means, the patch $x_k$ that gives the highest similarity score with the anchor patch $x^a$ is chosen from all patches in the batch, with the exception of the positive pair $x^+$ and the anchor patch $x^a$ itself (as depicted in Figure \ref{fig:sampling}). Overall, this approach underscores the significance of selecting suitable data augmentation techniques for \ac{hsi} and brings to the forefront the potential advantages of utilizing self-supervised contrastive learning to derive meaningful representations from \ac{hsi} data.
\begin{table*}[!htb]
	\centering
	\caption{\small{A comparison to popular hyperspectral datasets}. *}
	\label{tab:EnHyperSet-1}
	\scalebox{0.83}{
		\resizebox{\textwidth}{!}{%
			\begin{tabular}{@{}lccccccc@{}}
				\toprule
				Dataset &
				Number of Bands &
				Size &
				Spectral Range &
				Number of Images &
				Spatial Resolution &
				Imaging Location &
				Platform Type \\ 
				\midrule
				Indian Pines &
				200 &
				145x145 &
				400-2500nm &
				1 &
				30m &
				Indiana, USA &
				Airborne \\
				Pavia Centre &
				102 &
				1096x1096 &
				430-860nm &
				1 &
				\textbf{1.3m} &
				Pavia, Italy &
				Airborne \\
				Salinas &
				204 &
				512x217 &
				{\color[HTML]{FF0000} \textbf{360-2500nm}} &
				1 &
				3.7m &
				Salinas Valley, California, USA &
				Airborne \\
				Harvard &
				31 &
				1392x1040 &
				420-720nm &
				50 &
				- &
				Harvard, USA &
				Airborne \\
				Botswana &
				145 &
				1476x256 &
				400-2500nm &
				1 &
				30m &
				Botswana &
				Airborne \\
				Chikusei &
				100 &
				{\color[HTML]{FF0000} \textbf{2517x2335}} &
				263-1018nm &
				1 &
				2.5m &
				Chikusei, Japan &
				Airborne \\
				EnHyperSet-1 &
				{\color[HTML]{FF0000} \textbf{224}} &
				1276x1248 &
				420-2450nm &
				{\color[HTML]{FF0000} \textbf{800}} &
				30m &
				{\color[HTML]{FF0000} \textbf{Global, on demand}} &
				{\color[HTML]{FF0000} \textbf{Spaceborne}} \\
				\bottomrule
			\end{tabular}%
		}
	}
	
	\tiny{* Color Convention: {\color[HTML]{FF0000} \textbf{best}}}
\end{table*}
\section{Results}
\subsection{Experimental setup}
\subsubsection{Dataset}
\label{subsec:dataset}
The study employs a specialized dataset named EnHyperSet-1, tailored for hyperspectral deep learning applications. Sourced from the \ac{enmap} portal, the dataset includes 800 images of size $1,300$ by $1,200$ pixels, providing over 8,100 samples, segregated into distinct levels\cite{center2023enmap}: Level 1B, Level 1C, and Level 2A. Level 1B images present radiometrically corrected radiance measurements. Level 1C images are an orthorectified version of Level 1B, with geolocation assigned to each pixel. Meanwhile, Level 2A images build upon Level 1C by making atmospheric corrections for aspects like gas absorption and aerosol scattering and calculating the reflectance values for each pixel. Each high-resolution hyperspectral image pixel represents a 30m x 30m terrestrial section. The spectral resolution ranges from 6.5 nm to 12 nm, spanning from 420 nm to 2450 nm.

The EnHyperSet-1 dataset stands out for its expansive spectral range, fine spectral resolution, and optimal patch dimensions, as highlighted in Table \ref{tab:EnHyperSet-1}. The 160x160 pixel patches balance spatial detail capture and efficient processing. Featuring diverse scenes, including urban, forest, and agricultural settings, EnHyperSet-1, though recent, has demonstrated its efficacy in training algorithms, evident from HyperKon's performance when trained on it.

The HyperKon network was pretrained on EnHyperSet-1 using self-supervised contrastive learning with the NT-Xent loss. The training dataset comprised a diverse set of hyperspectral patches measuring 160 x 160 pixels, with a 5\% overlap from various scenes, including urban, agricultural, and natural landscapes. The network was trained for a total of 10,000 epochs with a batch size of 32, using the Adam optimizer with an initial learning rate of 1e-4 and a Stepped learning rate scheduler.

\subsubsection{Ablation Study}
\label{subsub:ablation}
After pretraining, the HyperKon network was evaluated using Top-K nearest neighbour retrieval accuracy to assess feature representation capabilities from hyperspectral data. Higher accuracy suggests successful capture of essential spectral and spatial hyperspectral characteristics, beneficial for tasks such as classification or segmentation.

\begin{figure}[!htb]
	\centering
	\includegraphics[width=0.4\textwidth]{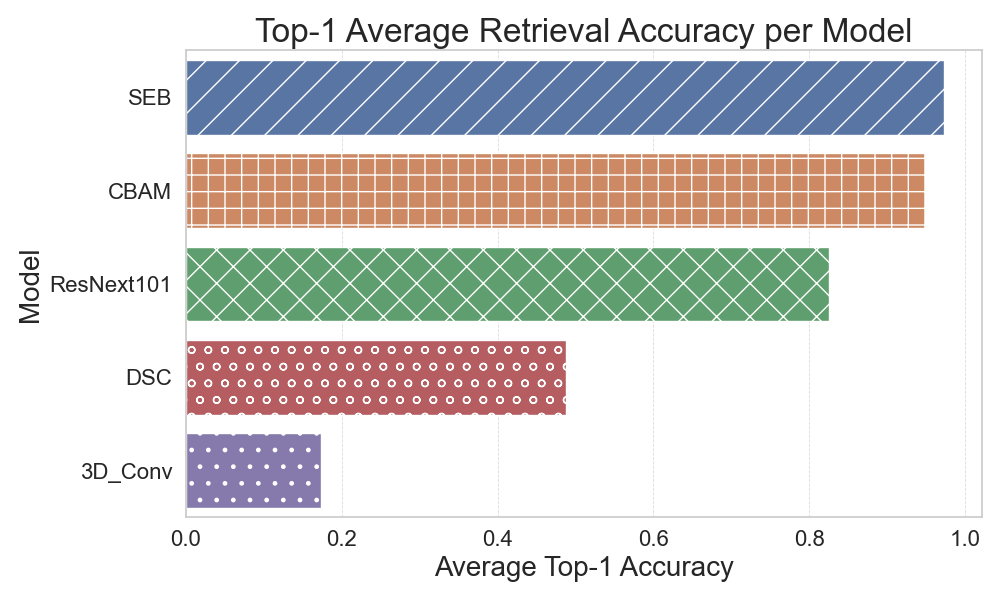}
	\caption{\label{fig:hyperkon_pretraining}
		\footnotesize{This showcases the top-1 HSI retrieval accuracy achieved by various versions of the HyperKon model during the pretraining phase. The performance of each version is presented as a bar in the chart, illustrating how the integration of different components, such as 3D convolutions, DSC, the CBAM, and the SEB, affected the model's accuracy. The chart underscores the superior performance of SEB.}
	}
\end{figure}

The results, as depicted in Figure \ref{fig:hyperkon_pretraining}, confirm the effectiveness of HyperKon in representing hyperspectral data features. This highlights its potential in hyperspectral image analysis and its role in enhancing the performance of deep learning models in areas like precision agriculture, environmental monitoring, and land cover classification. The findings also emphasize the importance of designing \ac{cnn} backbones specifically for hyperspectral data, which offer better performance than traditional RGB-Native backbones.

\subsubsection{Band Attention}
\begin{figure}[!htb]
	\centering
	\includegraphics[width=0.4\textwidth]{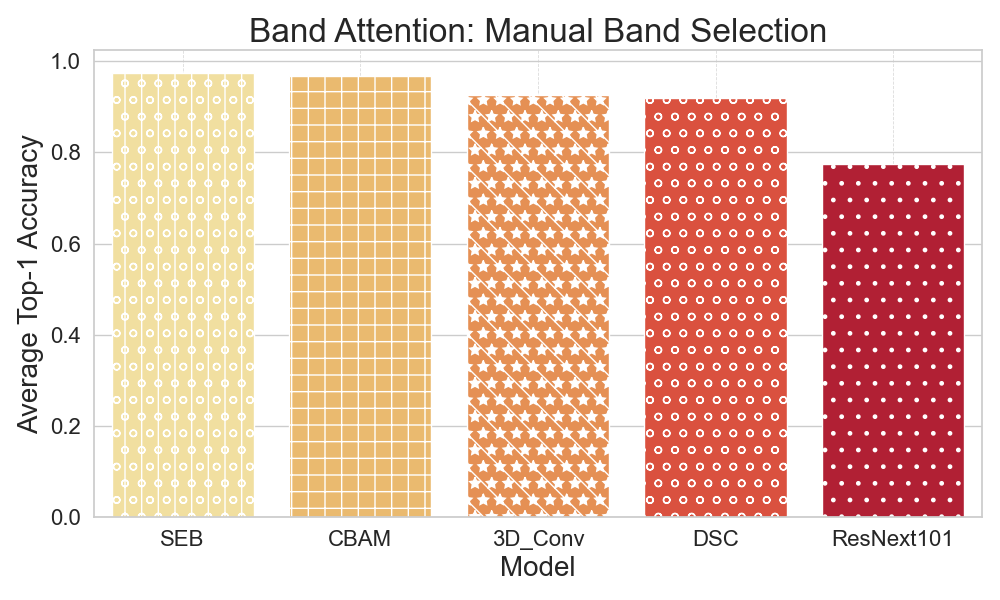}
	\caption{\label{fig:mbs_ba} 
		\footnotesize{Top-1 HSI retrieval accuracy for the 3D Conv, SEB, and CBAM models when manual band selection is employed. It shows an initial advantage for 3D Conv, but over time, the SEB and CBAM models catch up to similar levels of performance.}
	}
\end{figure}

\vspace{-2mm}
We conducted another study focusing on dimensionality reduction techniques which are commonly applied to this type of data \cite{li2018discriminant, kumar2020feature, zhang2020review}. \ac{pca}, Manual, and Full Band Selection were among the methods investigated. Dimensionality reduction is critical in hyperspectral image processing because it not only streamlines computation but also has the potential to refine the feature extraction or transfer learning process by discarding redundant information.
\begin{figure}[!htb]
	\centering
	\includegraphics[width=0.4\textwidth]{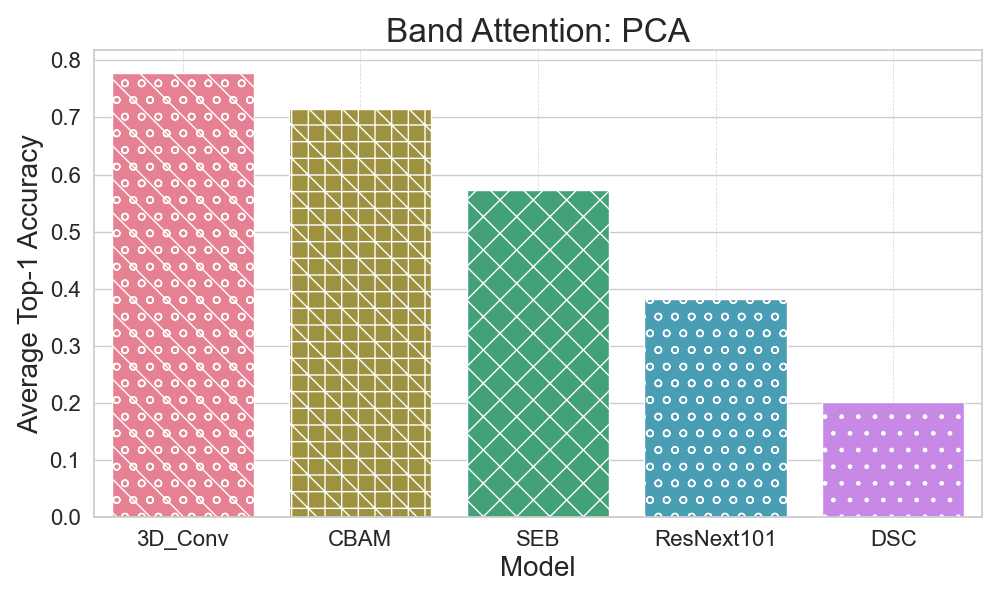}
	\caption{\label{fig:pca_ba} 
		\footnotesize{Top-1 HSI retrieval accuracy for the 3D Conv, SEB, and CBAM models following dimensionality reduction using PCA. The graph indicates a superior performance by the 3D Conv model.}
	}
\end{figure}
Contrary to our initial expectations, as shown in Figure \ref{fig:pca_ba}, the 3D Conv model outperformed both the \ac{seb} and \ac{cbam} models when dimensionality was reduced using \ac{pca}. \ac{seb} and \ac{cbam} generally demonstrate their proficiency by effectively selecting and prioritising information across the entire spectrum of hyperspectral bands; however, their advantage appears to diminish when \ac{pca} is used, potentially rendering their roles redundant. With the ability to attend to locally contiguous channels, 3D convolutions demonstrate superior performance when band selection is conducted \textit{a priori} using \ac{pca} with 112 components. As suggested in Figure \ref{fig:mbs_ba}, manual band selection gives 3D convolutions a distinct advantage. However, most notably, \ac{seb} and \ac{cbam} clearly surpass 3D convolution when no band selection is carried out. In these circumstances, 3D convolution is at a disadvantage due to its constraint of only attending to locally contiguous channels. This observation confirms our hypothesis: in the absence of \textit{a priori} band selection, attention mechanisms have the capacity to attend across the entire channel spectrum outperform those with more limited range.
Further analysis revealed that the \ac{seb} and \ac{cbam} model variants performed optimally without the application of \ac{pca}. See Figure \ref{fig:mbs_ba}. This clearly demonstrates their robustness and ability to manage the high dimensionality inherent in hyperspectral data.

\subsection{Super Resolution}

For the super-resolution evaluation, we utilized datasets such as \ac{enmap} and Pavia, each featuring different spectral bands. Specific bands from the EnHyperSet-1 dataset were selected to match the wavelength of each dataset, optimizing the training of HyperKon. This tailored approach aimed to enhance the super-resolution outcomes. The data preparation adhered to methodologies outlined in \cite{bandara2021hyperspectral} and \cite{zheng2020hyperspectral}.

The results from HyperKon were benchmarked against traditional RGB-trained backbones using metrics \cite{singh2014quality, deborah2015comprehensive, chaithra2018survey} such as \ac{cc}, \ac{sam}, \ac{rmse}, \ac{ergas}, and \ac{psnr}. This setup ensured a balanced comparison and highlighted the potential of HyperKon in a variety of hyperspectral tasks.
\vspace{-5mm}
\subsubsection{HyperKon as a Perceptual Loss Network}
We evaluated the use of HyperKon as a perceptual loss network for hyperspectral deep learning downstream tasks, comparing its performance to traditional RGB-trained backbones. See Table \ref{tab:Pavia_pansharpening} and \ref{tab:average_quantitative_pansharpening}. Specifically, we focused on tasks such as hyperspectral image super-resolution.
Our experiments demonstrated that using HyperKon as a perceptual loss network led to improved performance in these tasks compared to using RGB-trained backbones. This is primarily because HyperKon was specifically designed and trained for hyperspectral data, allowing it to capture more relevant features and better supervise the deep learning models in these tasks.

\begin{table}[!htb]
	\centering
	\caption{\small{Average quantitative results: Pavia Center\cite{plaza2009recent}}. *}
	\label{tab:Pavia_pansharpening}
	\scalebox{0.85}{
		\resizebox{\linewidth}{!}{%
			\begin{tabular}{@{}lccccc@{}}
				\toprule
				& \multicolumn{5}{c}{Pavia Center Dataset} \\ 
				\cmidrule{2-6} 
				Method & CC$\uparrow$ & SAM$\downarrow$ & RMSE$\downarrow$ & ERGAS$\downarrow$ & PSNR$\uparrow$ \\
				\midrule
				HySure\cite{simoes2014convex} & 0.966 & 6.13 & 1.8 & 3.77 & 35.91 \\
				HyperPNN\cite{he2019hyperpnn} & 0.967 & 6.09 & 1.67 & 3.82 & 36.7 \\
				PanNet\cite{yang2017pannet} & 0.968 & 6.36 & 1.83 & 3.89 & 35.61 \\
				Darn\cite{zheng2020hyperspectral} & 0.969 & 6.43 & 1.56 & 3.95 & 37.3 \\
				HyperKite\cite{bandara2021hyperspectral} & \textbf{0.98} & 5.61 & \textbf{1.29} & \textbf{2.85} & \textbf{38.65} \\
				SIPSA\cite{lee2021sipsa} & 0.948 & 5.27 & 2.38 & 4.52 & 33.65 \\
				GPPNN\cite{xu2021deep} & 0.963 & 6.52 & 1.91 & 4.05 & 35.36 \\
				HyperTransformer\cite{bandara2022hypertransformer} &
				{\color[HTML]{4472C4} \underline{0.9881}} &
				{\color[HTML]{4472C4} \underline{4.1494}} &
				{\color[HTML]{4472C4} \underline{0.9862}} &
				{\color[HTML]{4472C4} \underline{0.5346}} &
				{\color[HTML]{4472C4} \underline{40.9525}} \\
				Ours &
				{\color[HTML]{FF0000} \textbf{0.9883}} &
				{\color[HTML]{FF0000} \textbf{3.9551}} &
				{\color[HTML]{FF0000} \textbf{0.9369}} &
				{\color[HTML]{FF0000} \textbf{0.5152}} &
				{\color[HTML]{FF0000} \textbf{41.9808}} \\
				\bottomrule
			\end{tabular}%
		}
	}
	
	\tiny{* Color Convention: {\color[HTML]{FF0000} \textbf{first}}, {\color[HTML]{4472C4} \underline{second}}, {\textbf{third}}. RMSE values are $\times 10^{-2}$ }
\end{table}

\begin{table}[!htb]
	\centering
	\caption{\small{A comparison of the average quantitative pansharpening results}*}
	\label{tab:average_quantitative_pansharpening}
	\scalebox{0.95}{
		\resizebox{\columnwidth}{!}{%
			\begin{tabular}{@{}lccccccc@{}}
				\toprule
				& \multicolumn{2}{c}{\textbf{Botswana Dataset}} & \multicolumn{2}{c}{\textbf{Chikusei Dataset}} & \multicolumn{2}{c}{\textbf{Pavia Dataset}} \\
				\cmidrule(lr){2-3} \cmidrule(lr){4-5} \cmidrule(lr){6-7}
				\textbf{Metric} & \textbf{RGB-Native} & \textbf{HS-Native} & \textbf{RGB-Native} & \textbf{HS-Native} & \textbf{RGB-Native} & \textbf{HS-Native} \\
				\midrule
				CC$\uparrow$ & 0.9104 & {\color[HTML]{FF0000} \textbf{0.9411}} & {\color[HTML]{FF0000} \textbf{0.9801}} & 0.9777 & 0.9881 & {\color[HTML]{FF0000} \textbf{0.9883}} \\
				SAM$\downarrow$ & 3.1459 & {\color[HTML]{FF0000} \textbf{2.5798}} & {\color[HTML]{FF0000} \textbf{2.2547}} & 2.4192 & 4.1494 & {\color[HTML]{FF0000} \textbf{3.9551}} \\
				RMSE$\downarrow$ & 0.0233 & {\color[HTML]{FF0000} \textbf{0.0193}} & {\color[HTML]{FF0000} \textbf{0.0123}} & 0.0131 & 0.0098 & {\color[HTML]{FF0000} \textbf{0.0093}} \\
				ERGAS$\downarrow$ & 0.6753 & {\color[HTML]{FF0000} \textbf{0.5249}} & {\color[HTML]{FF0000} \textbf{0.8662}} & 0.9193 & 0.5346 & {\color[HTML]{FF0000} \textbf{0.5152}} \\
				PSNR$\uparrow$ & 27.3925 & {\color[HTML]{FF0000} \textbf{29.4128}} & {\color[HTML]{FF0000} \textbf{36.8861}} & 36.2889 & 40.9525 & {\color[HTML]{FF0000} \textbf{41.9808}} \\
				\bottomrule
			\end{tabular}%
		}
	}
	
	\tiny{* Color convention: {\color[HTML]{FF0000} \textbf{Best}}}
\end{table}

In summary, our experimental results showed that the HyperKon network, pretrained using self-supervised contrastive learning on \ac{enmap} \ac{hsi}, was highly effective as a hyperspectral native \ac{cnn} backbone and as a perceptual loss network for hyperspectral deep learning downstream tasks. These results highlight the potential of the HyperKon network for a wide range of applications in hyperspectral image analysis, paving the way for future research and development in this field
\begin{table}[!htb]
	\centering
	\caption{Average Classification accuracies of different methods on Hyperspectral Image Classification}
	\label{tab:hsic_10p}
	\scalebox{0.47}{
		\resizebox{\textwidth}{!}{%
			\begin{tabular}{@{}lcccccccc@{}}
				\toprule
				Datasets & Metrics & SSAN\cite{sun2019spectral} & SSRN\cite{zhong2017spectral} & RvT\cite{heo2021rethinking} & HiT\cite{yang2022hyperspectral} & SSFTT\cite{sun2022spectral} & QSSPN-3\cite{zhang2023quantum} & HyperKon \\ \midrule
				& OA(\%) & 89.46 & 91.85 & 83.85 & 90.59 & {\color[HTML]{4472C4} 96.35} & 95.87 & {\color[HTML]{FF0000} 98.77} \\ 
				IP& AA(\%) & 85.99 & 81.51 & 79.67 & 86.71 & 89.99 & {\color[HTML]{4472C4} 96.40} & {\color[HTML]{FF0000} 97.82} \\ 
				& Kappa(\%) & 88.04 & 90.73 & 81.68 & 89.27 & {\color[HTML]{4472C4} 95.82} & 95.34 & {\color[HTML]{FF0000} 98.60} \\ \midrule
				& OA(\%) & 99.15 & 99.63 & 97.37 & 99.43 & 99.52 & {\color[HTML]{4472C4} 99.71} & {\color[HTML]{FF0000} 99.89} \\ 
				PU& AA(\%) & 98.70 & 99.29 & 95.86 & 99.09 & 99.20 & {\color[HTML]{4472C4} 99.43} & {\color[HTML]{FF0000} 99.76} \\ 
				& Kappa(\%) & 98.87 & 99.51 & 96.52 & 99.24 & 99.36 & {\color[HTML]{4472C4} 99.61} & {\color[HTML]{FF0000} 99.86} \\ \midrule
				& OA(\%) & 98.92 & 99.31 & 98.11 & 99.38 & 99.53 & {\color[HTML]{4472C4} 99.66} & {\color[HTML]{FF0000} 99.99} \\ 
				SA& AA(\%) & 99.33 & 99.70 & {\color[HTML]{4472C4} 98.83} & 99.70 & 99.72 & 99.81 & {\color[HTML]{FF0000} 99.98} \\ 
				& Kappa(\%) & 98.80 & 99.23 & 97.90 & 99.31 & 99.47 & {\color[HTML]{4472C4} 99.63} & {\color[HTML]{FF0000} 99.99} \\ \bottomrule
			\end{tabular}%
		}
	}
	
	\tiny{* Color convention: {\color[HTML]{FF0000} \textbf{Best}}, {\color[HTML]{4472C4}  \textbf{2nd Best}}}
\end{table}

\subsection{Transfer Learning Capability of HyperKon}
In addition to the super-resolution task, the performance of the HyperKon network was evaluated on the hyperspectral image classification task using the Indian Pines, Pavia University, and Salinas Scene \cite{GREEN1998227} dataset. This dataset, covering a wide range of crops, serves as an excellent benchmark for assessing the accuracy and proficiency of the HyperKon network in classifying different crops. The frozen backbone of the HyperKon network, which had been pretrained on a variety of hyperspectral data, was utilized for this task

The results were juxtaposed with those of established methods such as SSAN \cite{sun2019spectral}, SSRN \cite{zhong2017spectral}, RvT \cite{heo2021rethinking}, HiT \cite{yang2022hyperspectral}, SSFTT \cite{sun2022spectral}, and QSSPN\cite{zhang2023quantum}. The assessment utilized overall accuracy (OA), average accuracy (AA), and Kappa coefficient (Kappa) metrics. As shown in Table \ref{tab:hsic_10p}, HyperKon consistently matched or surpassed other networks, highlighting its robustness and adaptability to new data. The commendable accuracy underscores the model's ability to harness discriminative features from its self-supervised learning phase, leading to accurate predictions in hyperspectral image classification.
\section{Discussion}
The comparative analysis in Table \ref{tab:average_quantitative_pansharpening} demonstrates HyperKon's proficiency in hyperspectral super-resolution, validated against traditional RGB models on datasets including \ac{enmap}, Botswana, Chikusei, and Pavia. Metrics like \ac{cc}, \ac{sam}, \ac{rmse}, \ac{ergas}, and \ac{psnr} were used for evaluation. HyperKon's design, which incorporates \ac{seb} \cite{hu2018squeeze}, is pivotal in its superior performance by enhancing band-specific information processing capabilities.

Although HyperKon showed exceptional results, its performance on the Chikusei dataset was not optimal compared to an RGB-Native model. This suggests potential improvements through training on a more diverse set of hyperspectral images to enhance its generalization capacity.

Additionally, HyperKon's capability for transfer learning was highlighted through its performance in hyperspectral classification tasks, as detailed in Table \ref{tab:hsic_10p}. Its self-supervised pretraining on hyperspectral data allows for effective feature extraction even with limited labeled data, making it a robust tool for applications where acquiring extensive labeled datasets is impractical or expensive.

\section{Conclusions}
In this study, we introduced HyperKon, a self-supervised contrastive network specifically tailored for hyperspectral image processing. Through rigorous training on a substantial dataset consisting of 8,180 patches of \acp{hsi}, HyperKon demonstrated exceptional performance in two critical downstream tasks: hyperspectral super-resolution and image classification, particularly in the context of crop mapping. Our findings highlight the significant potential of self-supervised contrastive networks in hyperspectral image processing and lay a robust foundation for future research in this rapidly expanding field.

 \paragraph{Acknowledgement:}
 The author's of this paper wish to extend their profound gratitude to \href{https://sixteensands.com/}{Sixteen Sands Ltd} for funding this research work, and the \href{https://www.enmap.org/}{German Aerospace Center (DLR)} for providing EnMAP free of charge.

\newpage
{\small
	\bibliographystyle{ieee_fullname}
	\bibliography{egbib}
}


\begin{acronym}[RANSAC] 

\acro{bmvc}[BMVC]{British Machine Vision Conference}
\acro{iccv}[ICCV]{International Conference on Computer Vision}

\acro{rmse}[RMSE]{Root Mean Square Error}
\acro{psnr}[PSNR]{Peak Signal-to-Noise Ratio}
\acro{cc}[CC]{Correlation Coefficient}
\acro{sam}[SAM]{Spectral Angle Mapper}
\acro{ergas}[ERGAS]{Erreur Relative Globale Adimensionnelle de Synthèse}
\acro{cnn}[CNN]{Convolutional Neural Network}
\acrodefplural{cnn}{Convolutional Neural Networks}
\acro{gan}[GAN]{Generative Adversarial Network}
\acro{fe}[FE]{Feature Extractor}
\acro{lfe}[LFE]{Learnable Feature Extractor}
\acro{dsc}[DSC]{Depthwise Separable Convolutions}
\acro{seb}[SEB]{Squeeze and Excitation Block}
\acro{cbam}[CBAM]{Convolutional Block Attention Module}
\acro{vit}[ViT]{Vision Transformers}
\acro{tnt}[TNT]{Transformer-iN-Transformer}
\acro{swint}[SwinT]{Swin Transformer}
\acro{3dswint}[3DSwinT]{3-D Swin Transformer}
\acro{gcn}[GCN]{Graph Convolutional Networks}
\acro{mrf}[MRF]{Markov Random Field}
\acro{capsnet}[CapsNet]{Capsule Networks}
\acro{gnn}[GNN]{Graph Neural Networks}
\acro{gcn}[GNN]{Graph Convolutional Networks}
\acro{mlgcn}[MLGCN]{Mutual Learning Graph Convolutional Network}
\acro{awgcn}[AwGCN]{Attention-weighted Graph Convolutional Networks}
\acro{pca}[PCA]{Principal Component Analysis}
\acro{vae}[VAE]{Variational Autoencoder}
\acro{mhfsa}[MHFSA]{Multi-Head Feature Soft Attention}
\acro{mha}[MHA]{Multi-Head Attention}
\acro{bn}[BN]{Batch Normalisation}

\acro{soc}[SOC]{Soil Organic Carbon}
\acro{uav}[UAV]{Unmanned Aerial Vehicle}
\acrodefplural{uav}{Unmanned Aerial Vehicles}
\acro{hsr}[HSR]{Hyperspectral Super-Resolution}
\acro{dss}[DSS]{Decision Support System}
\acro{rnn}[RNN]{Recurrent Neural Networks}
\acro{rn}[ResNet]{Residual Network}
\acro{nce}[NCE]{Noise Contrastive Estimation}
\acro{hsi}[HSI]{HyperSpectral Image}
\acro{pan}[PAN]{Panchromatic}
\acro{lr-hsi}[LR-HSI]{Low-Resolution HyperSpectral Image}

\acro{nt-xent}[NT-Xent]{Normalized Temperature-Scaled Cross Entropy}
\acro{hspl}[HSPL]{HyperSpectral Perceptual Loss}

\acro{dlr}[DLR]{German Aerospace Center}
\acro{esa}[ESA]{European Space Agency}
\acro{enmap}[EnMAP]{Environmental Mapping and Analysis Program}
\acro{chime}[CHIME]{Copernicus Hyperspectral Imaging Mission for the Environment}

\end{acronym}

\end{document}